\documentclass[runningheads]{llncs}
\usepackage{graphicx}
\usepackage{cite}
\usepackage{amsmath,amssymb,amsfonts}
\usepackage{algorithmic}
\usepackage{graphicx}
\usepackage{textcomp}
\usepackage{array, makecell}

\usepackage{amsmath, amssymb,bm}
\usepackage{booktabs}
\usepackage{graphicx,epstopdf,subfigure}
\usepackage{multirow}
\usepackage{tabularx,threeparttable}
\usepackage{array}
\usepackage{xcolor}
\usepackage{comment}
\usepackage{mathtools}
\usepackage{bbm}
\usepackage[ruled]{algorithm2e}

\usepackage{url}  

\usepackage[colorlinks=true,citecolor=blue,urlcolor=blue]{hyperref}

\usepackage{upgreek}

\usepackage{fancyhdr}

\begin{document}
\title{Source-Free Domain Adaptive Fundus Image Segmentation with Denoised Pseudo-Labeling}
\titlerunning{Source-Free Domain Adaptive Segmentation}
%

\author{Cheng Chen, Quande Liu, Yueming Jin, Qi Dou, and Pheng-Ann Heng}
%
\authorrunning{C. Chen et al.}

%
\institute{Department of Computer Science and Engineering, \\The Chinese University of Hong Kong, Hong Kong, China}

\maketitle              

\begin{abstract}
	Domain adaptation typically requires to access source domain data to utilize their distribution information for domain alignment with the target data. However, in many real-world scenarios, the source data may not be accessible during the model adaptation in the target domain due to privacy issue.
	This paper studies the practical yet challenging source-free unsupervised domain adaptation problem, in which only an existing source model and the unlabeled target data are available for model adaptation.
	We present a novel denoised pseudo-labeling method for this problem, which effectively makes use of the source model and unlabeled target data to promote model self-adaptation from pseudo labels. 
	Importantly, considering that the pseudo labels generated from source model are inevitably noisy due to domain shift, we further introduce two complementary pixel-level and class-level denoising schemes with uncertainty estimation and prototype estimation to reduce noisy pseudo labels and select reliable ones to enhance the pseudo-labeling efficacy.
	Experimental results on cross-domain fundus image segmentation show that without using any source images or altering source training, our approach achieves comparable or even higher performance than state-of-the-art source-dependent unsupervised domain adaptation methods\footnote{Code is available at \url{https://github.com/cchen-cc/SFDA-DPL}}.
	
\end{abstract}

\section{Introduction}
Deep neural networks remain notoriously vulnerable to the domain shift between the training (source) and testing (target) data acquired under different conditions~\cite{gibson2018inter,ghafoorian2017transfer,kamnitsas2017unsupervised}.  
To address the domain shift problem, unsupervised domain adaptation (UDA) has been an active research topic. Existing UDA methods typically require to simultaneously access the source and target data for distribution alignment~\cite{zhang2018task,varsavsky2020test,chen2020unsupervised,xing2019adversarial,ju2020leveraging,zhao2018supervised,huo2018synseg,perone2019unsupervised,zhang2018translating,zhang2018multi}.  
However, in real-world scenarios, the source data are often inaccessible during model adaptation in the target domain, because medical data are strictly regulated and prohibitive to be shared before taking complex ethical procedures. 
An interesting yet less explored problem is: Can a model adapt to target domains without using the source data? 
This problem has high practical value as often only a learned model is given instead of the source training data due to data privacy and security issues. 

We are concerned with a more challenging \textit{source-free unsupervised domain adaptation} setting, in which only a well-trained source model and unlabeled target domain data are provided for model adaptation.
Such a setting has wider applicability than vanilla UDA, since data transmission is not required and the model adaptation is only performed in the target domain. 
Previous UDA methods relying on distribution alignment are not suitable in this scenario as the source distribution information are not accessible.
For example, the adversarial learning based methods~\cite{vu2019advent,kamnitsas2017unsupervised} need to align the source and target features within a single model, which is infeasible without source data. 
A few very recent studies have started to explore the possibility of adapting a model in the absence of source data. 
Bateson et al.~\cite{bateson2020source} introduce an auxiliary branch in source model to estimate class-ratio prior and propose a prior-regularized entropy minimization strategy for source-relaxed adaption.
Stan et al.~\cite{stan2021privacy} encode the source samples into a prototypical distribution, which is transferred to the target domain for distribution alignment. 
In \cite{karani2021test} and \cite{he2020self}, auto-encoders are pre-trained in the source domain for denoising or image reconstruction and are leveraged in target domain for adaptation.
However, all these methods need to alter the source-domain training with auxiliary branches or additional training tasks, hence cannot adapt the existing models that have already been well-trained on the source domains.  

Given only an existing model trained on the source domain and unlabeled target data, the challenge lies in how to realize the model self-adaptation using the knowledge purely from the target domain.
Our insight is to make the best use of the source model and target data via self-training~\cite{lee2013pseudo}, that is generating pseudo labels in the target domain with the model's own predictions and re-training the model based on the generated pseudo labels.
In real-world model deployment, the most common situation is the model adaptation cross different clinical centers, where the encountered domain shift caused by the scanner difference is generally minor compared with cross-modality discrepancy (e.g., CT and MRI). 
Considering this practical situation, a model's own predictions on target data can be reasonable and provide a meaningful supervision base for model self-training.
However, some inevitable noise, that is the incorrect predictions caused by domain shift, shall harm pseudo-labeling for model self-adaptation. How to reduce the noisy pseudo labels is crucial for yielding its best efficacy.

In this paper, we present a novel \textbf{d}enoised \textbf{p}seudo-\textbf{l}abeling (DPL) method for source-free unsupervised domain adaptation.
Our method roots in the pseudo-labeling strategy, which enables effective self-training in the target domain with only a pre-trained source model and unannotated target data, without using any source-domain data or altering source-domain training.
Importantly, to reduce the training noise in pseudo-labeling, we propose two complementary pixel-level and class-level denoising schemes with uncertainty estimation and  prototype estimation, to provide more discriminative and less noisy supervision for model adaptation.
First, we estimate the model's pixel-level prediction uncertainty, and identify potential unreliable pseudo labels with high uncertainty.
Second, we estimate the class-level prototypes motivated by the prototypical networks~\cite{snell2017prototypical} and calculate relative feature distance, to suggest noisy pseudo labels which lie further away from their corresponding class prototypes. 
We evaluate our method on two public fundus image datasets for the multi-label optic disc and cup segmentation, which are popular benchmarks for UDA tasks. 
Without requiring any source data, our method achieves comparable or even higher performance than state-of-the-art source-dependent UDA methods. 

\section{Method}
Fig. 1 illustrates our source-free UDA framework via denoised pseudo-labeling. In this section, we first present the pseudo-label learning for source-free UDA. Next, we propose two complementary \textit{pixel-level} and \textit{class-level} pseudo label denoising schemes. The training procedures are finally described.

\subsection{Pseudo-Labeling for Source-free UDA}
In source-free UDA setting, we are given a model $f^s:\mathcal{X}^s\to \mathcal{Y}^s$ trained from unknown source domain $ \mathcal{D}^s=(\mathcal{X}^s, \mathcal{Y}^s)$, and an unlabeled dataset $\{x_i^t\}_{i=1}^{N^t}$ from target domain $\mathcal{D}^t$, where $x_i^t\in \mathcal{X}^t$. 
The goal of source-free UDA is to adapt the model $f^s$ with only $\{x_i^t\}_{i=1}^{N^t}$, such that the obtained model $f^{s\rightarrow t}$ can perform well on the target domain distribution.
We consider fundus image segmentation, which is a multi-label segmentation problem with $x_i \in \mathbb{R}^{H\times W \times 3}$ and $y_i \in \{0, 1\}^{H\times W \times C}$ ($C$ denotes the class number).

Ideal model adaptation towards target distribution would require the ground truth labels at target domain, which are not available in the source-free UDA setting. To address this problem, we devise an effective approach to assign pseudo labels to the target samples so that the model could be adapted via supervised learning on the target domains. 
Specifically, given an unlabeled image from the target domain, we denote $p_v$ as its prediction probability from the source model on $v$-th pixel. Then the corresponding pseudo label can be generated as:
\begin{equation}
\hat{y}^t_v = \mathbbm{1}[ ~p_v\ge \gamma~ ],
\label{eq:pseudo}
\end{equation}
where $\mathbbm{1}(\cdot)$ is the indicator function, $\gamma\in(0,1)$ is a probability threshold to generate binary pseudo labels for the multi-label segmentation task. 
With the generated pseudo labels, we can then  adapt the source model towards target data distribution with supervised learning:
\begin{equation}
\mathcal{L}_{ce} = -\sum_{v}~[\hat{y}^t_v\cdot log(p_v)+(1-\hat{y}^t_v)\cdot log(1-p_v)].
\label{eq:celoss}
\end{equation}
The target function $f^{s\rightarrow t}$ is obtained by updating the source model over all target domain data $\{x_i^t\}_{i=1}^{N^t}$ with generated pseudo labels$\{\hat{y}_i^t\}_{i=1}^{N^t}$.

\begin{figure*}[!t]
	\centering
	\includegraphics[width=1.0\textwidth]{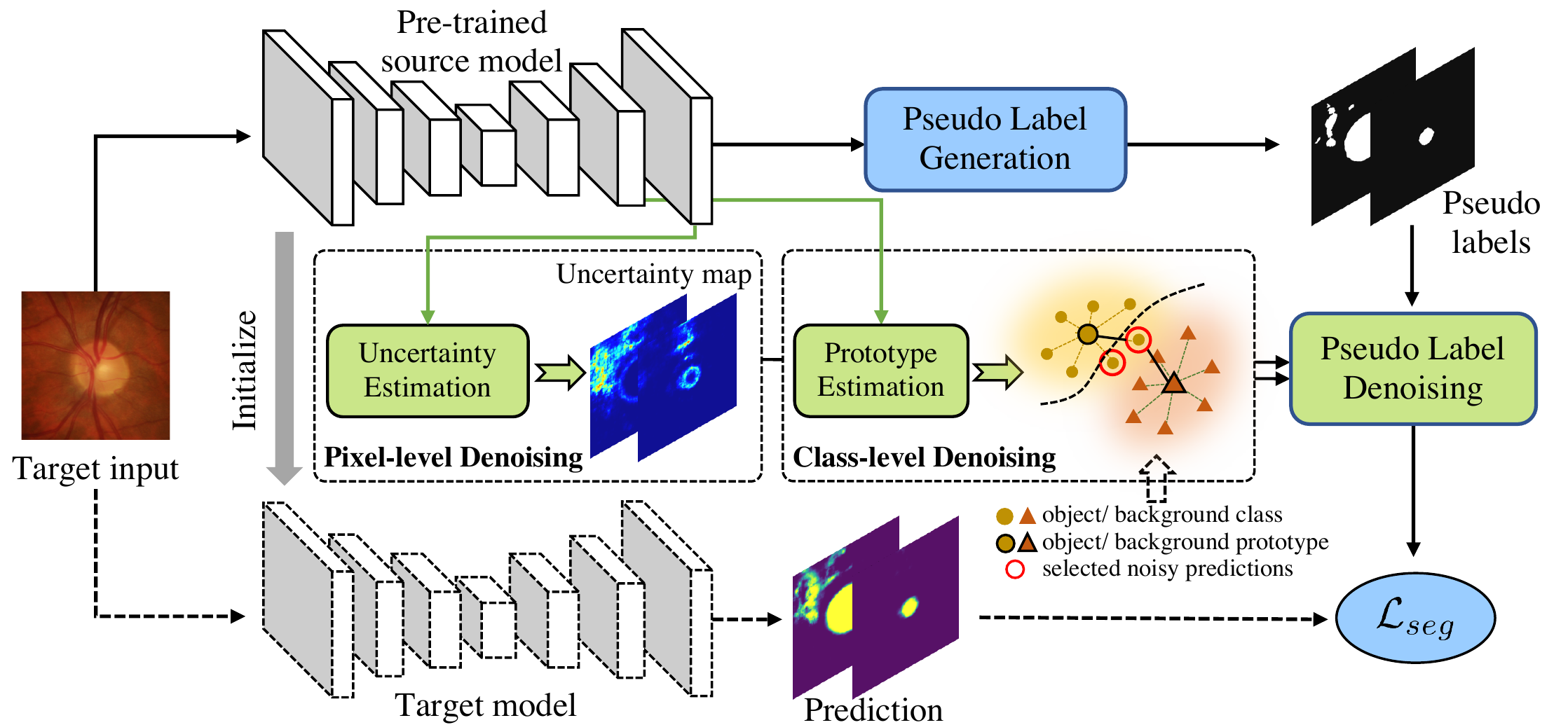}
	\caption{Overview of our proposed \textit{denoised pseudo-labeling} (DPL) for source-free UDA. Pseudo labels are generated from source model's predictions to guide the self-training in target domain. Two complementary denoising schemes with pixel-level uncertainty estimation and class-level prototype estimation are developed to reduce the noisy pseudo labels and preserve reliable ones to provide more helpful and less noisy supervision.}
	
\end{figure*}
\subsection{Pixel-level Denoising via Uncertainty Estimation}
Due to the discrepancy between source and target domains, the pseudo labels assigned from source model would unavoidably contain noisy and incorrect predictions. Adapting the model with false pseudo labels is harmful to further improve the model's discrimination capacity in target domain due to the accumulated segmentation errors.
To this end, we aim to carefully filter out the unreliable pseudo labels and only preserve the accurate ones to enhance the model adaptation in target domain.
The conventional way for pseudo label selection is directly based on the network output confidence probability to select high-confidence pseudo labels.
However, the model's predictions under domain shift are prone to be over-confident and incorrect predictions can also have high confidence scores~\cite{guo2017calibration}.
Motivated by the observation in previous supervised segmentation work that uncertainty measures correlate with incorrect predictions, we leverage the pixel-level uncertainty information from the model's predictions to better indicate the unreliability of pseudo label on each pixel of the segmentation map.

We estimate uncertainty with Monte Carlo Dropout~\cite{gal2016dropout} for Bayesian approximation. 
Specifically, given a target image $x^t$,
we enable dropout at inference and perform $K$ stochastic forward passes through the source model to obtain a set of predictive outputs, i.e., $p_k=f^s(x^t), k=1,...,K$. The uncertainty map $u^{H\times W \times C}$ is then estimated as the standard deviation of the $K$ outputs, i.e., $u=std(p_1,...,p_K)$. To utilize the uncertainty map to filter noisy pseudo labels, we define a label selection mask with a binary variable $m\subseteq\{0,1\}^{H\times W \times C}$, whose value at pixel $v$ is determined by the uncertainty measure $u_v$ with:
\begin{equation}
m_v = \mathbbm{1}[~u_v<\eta~],
\label{eq:mask}
\end{equation}
where $\eta$ is the uncertainty threshold. If uncertainty is sufficiently low, we have $m_v=1$ to select the corresponding pseudo label $\hat{y}^t_v$; otherwise, the pseudo label $\hat{y}^t_v$ is regarded as a noisy one and excluded from the loss calculation in Eq.~\ref{eq:celoss}.

\subsection{Class-level Denoising via Prototype Estimation}
The pixel-level denoising scheme estimates the prediction uncertainty for each pixel separately. 
However, the segmentation problem has structural characteristics and the segmentation regions of the same category are often highly correlated.  
In this regard, we also propose a new class-level denoising scheme by leveraging prototypes, i.e., class-wise feature centroids, to provide guidance for identifying potential unreliable pseudo labels.
This is based on the assumption that for a reliable pseudo label, its feature should lie closer to its corresponding class prototype, otherwise a potential noisy pseudo label is implied. 

Specifically, we exploit the relative feature distances to the class prototypes to determine whether a pseudo label is reliable. 
Given a target image $x^t$ and its generated pseudo label $\hat{y}^t$, we denote its feature map from the layer before the last convolution as $e_l\in \mathbb{R}^{H_l\times W_l\times L}$ ($L$ denotes the channel number), which is then bilinearly interpolated to $e\in \mathbb{R}^{H\times W\times L}$ to keep consistency with the dimensions of $\hat{y}^t\in \mathbb{R}^{H\times W\times C}$.
To extract the class-wise prototypes, we utilize the proposed uncertainty-guided pseudo labels to indicate the category and collect corresponding feature vectors for each object (foreground) class and background class, given that the ground truth of target images is unavailable. Herein, the binary object mask $b^{obj}\in \mathbb{R}^{H\times W\times C}$ can be obtained from $\hat{y}^t$ as $b^{obj}=\mathbbm{1}[\hat{y}^t=1]\mathbbm{1}[u<\eta]$, and the binary background mask $b^{
	bg}\in \mathbb{R}^{H\times W\times C}$ can be obtained as $b^{bg}=\mathbbm{1}[\hat{y}^t=0]\mathbbm{1}[u<\eta]$. Then, the class-wise prototypes $z^{obj}$ and $z^{bg}$, and the relative feature distance $d_v^{obj}$ and $d_v^{bg}$ between each feature vector for pixel $v$ to the two prototypes  are calculated as:
\begin{equation}
\small
z^{obj}=\frac{\sum_{v}e_v\cdot b^{obj}_v \cdot p_v}{\sum_{v}b^{obj}_v \cdot p_v},\\ ~~z^{bg}=\frac{\sum_{v}e_v\cdot b^{bg}_v\cdot (1-p_v)}{\sum_{v}b^{bg}_v \cdot (1-p_v)},
\end{equation}
\begin{equation}
\small
d^{obj}_v = ||e_v-z^{obj}||_2, ~~d^{bg}_v = ||e_v-z^{bg}||_2,
\end{equation}
in which we also incorporate the network output probability $p_v$ to weigh the contribution of each pixel to the prototype calculation. 
Based on the estimated prototypes and feature distance information, if $\hat{y}^t_v$ classifies pixel $v$ as an object, but its feature vector $e_v$ is further away from the object prototype $z^{obj}$ than the background prototype $z^{bg}$, it is likely that the pseudo label is unreliable. We then remove it from the selection mask to denoise the pseudo labels.
Based on this denoising scheme, the label selection mask in Eq.~\ref{eq:mask} is updated as:
\begin{equation}
\small
m_v = \mathbbm{1}[u_v<\eta]\mathbbm{1}[\hat{y}^t_v=1]\mathbbm{1}[d^{obj}_v<d^{bg}_v]+\mathbbm{1}[u_v<\eta]\mathbbm{1}[\hat{y}^t_v=0]\mathbbm{1}[d^{obj}_v>d^{bg}_v].
\label{eq:maskcombine}
\end{equation}
By combining the uncertainty estimation and prototype estimation in Eq.~\ref{eq:maskcombine}, a pseudo label $\hat{y}^t_v=1$ ($\hat{y}^t_v=0$) is selected when network's uncertainty on prediction is low and encoded feature lies closer to the object (background) prototype than the background (object) prototype. The segmentation loss with pseudo labels in Eq.~\ref{eq:celoss} is updated with label selection mask as $\mathcal{L}_{seg} = \sum_{v}~m_v*{L}_{ce,v}$.

\subsection{Training Procedures and Implementation Details}
Given the target images $\{{x^t_i}\}_{i=1}^{N^t}$ and the source model $f^s$, we first apply $f^s$ on all the target images to generate predictions. Pseudo labels are obtained from these predictions with Eq.~\ref{eq:pseudo} and denoised with Eq.~\ref{eq:maskcombine}. We then initialize the target model $f^{s\to t}$ with $f^s$ and optimize $f^{s\to t}$ with the denoised pseudo labels. 
We employ a MobileNetV2 adapted DeepLabv3+~\cite{chen2018encoder} as network backbone, following the previous work \cite{wang2019boundary} without the boundary branch. 
For uncertainty estimation with Monte Carlo Dropout, the dropout rate is set to 0.5 and 10 stochastic forward passes are performed to obtain the standard deviation of the output probabilities. 
The threshold $\gamma$ for multi-label task is set as 0.75 referring to \cite{wang2019boundary} and the uncertainty threshold $\eta$ is set as 0.05 by visually inspecting the uncertainty map. 
We use weak augmentations including Gaussian noise, contrast adjustment, and random erasing to slightly disturb the inputs when training $f^{s\to t}$, in order to make the predictions deviate from pseudo labels. 
The model is trained using Adam optimizer with momentum of 0.9 and 0.99, and learning rate of 2e-3. We train the target model with batch size 8 and 2 epochs. 
The framework is implemented with Pytorch 0.4.1 using one NVIDIA TitanXp GPU. 

\section{Experiments}
\subsubsection{Dataset and evaluation metrics.}
We validate our method on optic disc and cup segmentation for retinal fundus images, using datasets collected from different clinical centers with distribution shifts. Specifically, we employ the training set of the REFUGE challenge~\cite{orlando2020refuge} as the source domain, and the public RIM-ONE-r3~\cite{fumero2011rim} and Drishti-GS~\cite{sivaswamy2015comprehensive} datasets as two different target domains.
The source domain includes 400 annotated training images, and the two target domain data are split to 99/60 and 50/51 images for training/testing respectively, following the same experimental setup in \cite{wang2019boundary}.
Each image is pre-processed by cropping a 512$\times$512 disc region as the network input. 
For evaluation, we employ two commonly-used metrics, including the Dice coefficient for pixel-wise accuracy measure and the Average Surface Distance (ASD) for boundary agreement assessment. The higher Dice and lower ASD  indicate a better performance.

\subsubsection{Comparison with state-of-the-arts.}
We compare our method with recent state-of-the-art domain adaptation methods, including \textbf{BEAL}~\cite{wang2019boundary}: an UDA method with adversarial learning between \textit{source} and target data with boundary prediction, which is the best reported UDA model on cross-domain fundus image segmentation; \textbf{AdvEnt}~\cite{vu2019advent}: a popular UDA benchmark approach that encourages entropy consistency between the \textit{source} and target domains; \textbf{SRDA}~\cite{bateson2020source}: a domain adaptation approach that utilizes the task prior trained from \textit{source domain} for model adaptation in the target domain; \textbf{DAE}~\cite{karani2021test}: a domain adaptation approach which utilizes a denoising auto-encoder learned from \textit{source domain} to gradually refine the segmentation mask in target domain.
Note that these methods can serve as strong baselines in our source-free UDA setting since they utilize more source domain information by either accessing source domain data or altering source domain training, while our method is completely free of the source domain. 
Since we follow the same network backbone and data split as BEAL, we referenced the performance reported in their paper for comparison.  

\begin{table*}[!tbp]
	\centering
	\caption{Quantitative comparison of different methods on the target domain datasets. (Note: S. denotes source domain, - means the results are not reported by that methods.)
	}
	\begin{center}
		\setlength\tabcolsep{3.0pt}
		\resizebox{1.0\textwidth}{!}{%
			\begin{tabular}{l|cc|c|c|c|c}
				\toprule[1.5pt]
				
				\multirow{2}{*}{Methods} &\multicolumn{2}{c|}{Access/Alter}&\multicolumn{2}{c|}{Optic Disc Segmentation}&\multicolumn{2}{c}{Optic Cup Segmentation} \\
				\cline{2-7}
				{} &S. Data &S. Train &Dice[\%]&ASD[pixel]&Dice[\%]&ASD[pixel]\\
				\cline{1-7}
				\multicolumn{7}{c}{\textbf{RIM-ONE-r3}}\\
				\hline
				W/o adaptation &&&83.18$\pm${6.46} &24.15$\pm${15.58}&74.51$\pm${16.40}&14.44$\pm${11.27}\\
				Oracle~\cite{wang2019boundary} &&&96.80&--&85.60&--\\
				\hline
				\hline
				BEAL~\cite{wang2019boundary} &$\checkmark$&&89.80&--&\textbf{81.00}&--\\
				AdvEnt~\cite{vu2019advent}&$\checkmark$&&89.73$\pm${3.66}&9.84$\pm${3.86}&77.99$\pm${21.08}&\textbf{7.57$\pm${4.24}}\\
				
				SRDA~\cite{bateson2020source}&&$\checkmark$&89.37$\pm${2.70}&9.91$\pm${2.45}&77.61$\pm${13.58} &10.15$\pm${5.75}\\
				DAE~\cite{karani2021test}&&$\checkmark$&89.08$\pm${3.32}&11.63$\pm${6.84}&79.01$\pm${12.82}&10.31$\pm${8.45}\\
				
				\hline
				\textbf{DPL~(ours)}&&&\textbf{90.13$\pm${3.06}}&\textbf{9.43$\pm${3.46}}&79.78$\pm${11.05}&9.01$\pm${5.59}\\
				\hline
				\multicolumn{7}{c}{\textbf{Drishti-GS}}\\
				\hline
				W/o adaptation &&&93.84$\pm${2.91}&9.05$\pm${7.50}&83.36$\pm${11.95}&11.39$\pm${6.30}\\
				
				Oracle~\cite{wang2019boundary} &&&97.40&--&90.10&--\\
				\hline
				\hline
				BEAL~\cite{wang2019boundary} &$\checkmark$&&96.10&--&\textbf{86.20}&--\\
				AdvEnt~\cite{vu2019advent}&$\checkmark$&&96.16$\pm${1.65}&4.36$\pm${1.83}&82.75$\pm${11.08}&\textbf{11.36$\pm${7.22}}\\
				SRDA~\cite{bateson2020source}&&$\checkmark$&96.22$\pm${1.30}&4.88$\pm${3.47}&80.67$\pm${11.78}&13.12$\pm${6.48}\\
				DAE~\cite{karani2021test}&&$\checkmark$&94.04$\pm${2.85}&8.79$\pm${7.45}&83.11$\pm${11.89}&11.56$\pm${6.32}\\
				
				\hline
				\textbf{DPL~(ours)}&&&\textbf{96.39$\pm${1.33}}&\textbf{4.08$\pm${1.49}}&83.53$\pm${17.80}&11.39$\pm${10.18}\\
				
				\hline
				\toprule[1.5pt]

		\end{tabular}}
	\end{center}
	%
\end{table*}

\begin{figure*}[!t]
	\centering
	\includegraphics[width=0.9\textwidth]{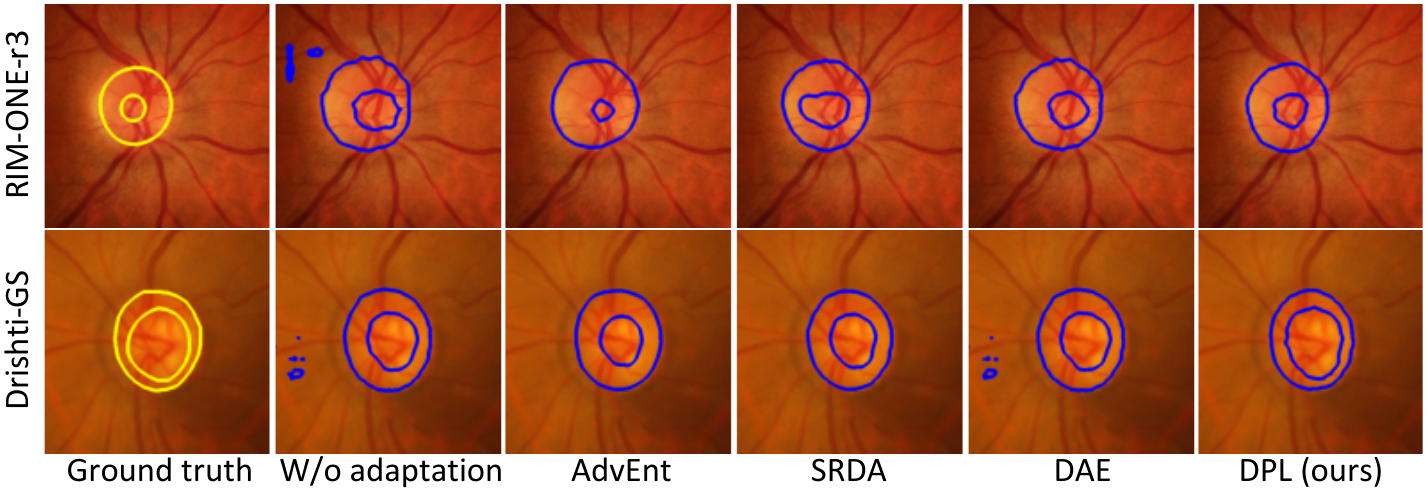}
	\caption{Qualitative comparison of adaptation performance with different methods.}
	\label{fig:compare}
\end{figure*}
The quantitative comparison results are presented in Table 1, where we also include the ``\textbf{W/o adaptation}" lower bound and the supervised training upper bound in target domain (referred as ``\textbf{Oracle}" with results from \cite{wang2019boundary}). As observed, different domain adaptation methods can generally improve the performance over baseline, with BEAL which is the current state-of-the-art UDA method in fundus image segmentation performing the best. This benefits from their joint utilization of source and target domain data to perform adversarial alignment in both boundary and predicted entropy.
Notably, without using any source data during adaptation, our method achieves 90.13\% and 96.39\% Dice scores for optic disc segmentation on the two target datasets respectively, which are even higher than the BEAL method. 
This indicates that source-free UDA would not necessarily underperform the vanilla UDA. A possible reason is that vanilla UDA heavily relies on finding an invariant latent space between source and target distributions which could be difficult, while our source-free method directly adapts the model towards target domain hence can capture more discriminative representations at the target distribution.
In addition, without altering source training, our method also presents clear improvements over SRDA and DAE on both target domains.
These results affirm the superiority of our method to adapt the model without using any source domain information, thanks to the proposed  denoising schemes to explicitly filter out unreliable pseudo labels to facilitate the discriminative learning from target samples.
\begin{figure*}[!t]
	\centering
	\includegraphics[width=1.0\textwidth]{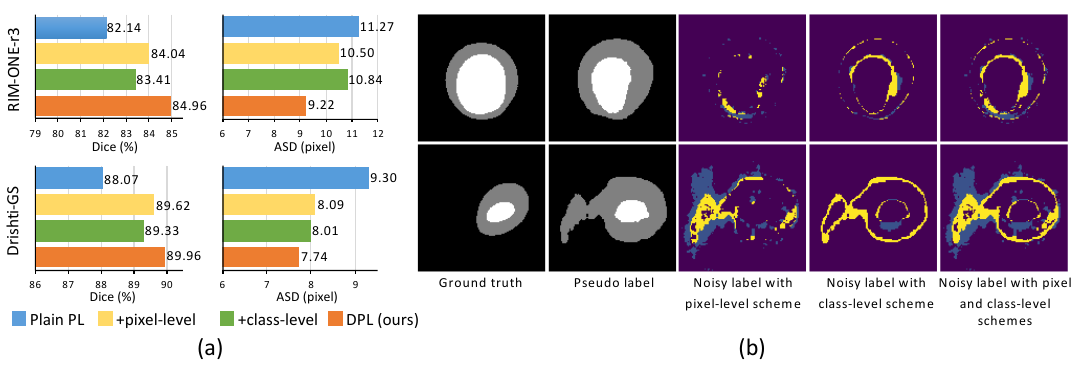}
	\caption{(a) Ablation results with different denoising schemes in our method; (b) Examples of noisy pseudo labels identified with different denoising schemes. The correctly and falsely identified noisy pseudo labels are indicated in yellow and blue colors.}
	\label{fig:ablation}
\end{figure*}
\subsubsection{Ablation study.}
We conduct ablation analysis to investigate the two denoising schemes. 
Fig.~\ref{fig:ablation} (a) shows that adding the pixel-level or class-level denoising scheme can both clearly improve the adaptation performance in terms of Dice score and ASD value over plain pseudo labeling. Besides, integrating the two complementary denoising schemes completes our method and yields further improvements on both two target domains. 
Moreover, Fig.~\ref{fig:ablation} (b) shows that each denoising scheme could identify different subsets of noisy pseudo labels and combining them could complement each other. 
Precisely identifying all the noisy pseudo labels is indeed challenging and some misidentification could exist as observed in Fig.~\ref{fig:ablation} (b). We consider this is tolerable as it is more critical to ensure the remaining pseudo labels are reliable for model self-training, thus the sensitivity of finding noisy pseudo labels is more important than the specificity.

Table~\ref{tab:placcuracy} shows the accuracy of the selected pseudo labels for quantitative comparison. We see that both two denoising schemes improve the pseudo label accuracy over the plain pseudo-labeling strategy, and their combination could further increase the accuracy on both optic disc and cup in the two target domains.
It is worth noting that the proposed denoising schemes work much more effective when being used to denoise the pseudo labels than being used as post-processing for the segmentation results. 
We tried to apply the denoising schemes to post-process the segmentation results and found that the performance was even lower than the w/o adaptation model. The reason could be that denoising pseudo labels has a certain tolerance of noisy label misidentification, while such errors would directly decrease the performance if taking as post-processing. 

\section{Conclusion}
\begin{table*}[!t]
	\centering
	
	\caption{Comparison of pseudo-label selection accuracy with different methods.
	}
	\begin{center}
		\setlength\tabcolsep{3.0pt}
		\resizebox{0.8\textwidth}{!}{%
			\begin{tabular}{l|>{\centering}p{2.2cm}|>{\centering}p{2.2cm}|>{\centering}p{2.2cm}|>{\centering\arraybackslash}p{2.2cm}}
				\toprule[1.5pt]
				\multirow{2}{*}{Methods} &\multicolumn{2}{c|}{\textbf{RIM-ONE-r3}}&\multicolumn{2}{c}{\textbf{Drishti-GS}}\\
				\cline{2-5}
				
				{} &\textit{PL Acc.\textsubscript{disc}}[\%] &\textit{PL Acc.\textsubscript{cup}}[\%]  &\textit{PL Acc.\textsubscript{disc}}[\%] &\textit{PL Acc.\textsubscript{cup}}[\%]  \\
				
				\cline{1-5}
				
				Plain PL&71.39&68.09&90.60&95.39\\
				+pixel-level &74.19&70.20&90.88&95.60\\
				+class-level &80.80&70.92&93.75&95.48\\
				DPL (ours) &\textbf{81.35}&\textbf{72.26}&\textbf{93.88}&\textbf{95.69}\\
				
				\toprule[1.5pt]

		\end{tabular}}
	\end{center}
	\label{tab:placcuracy}
\end{table*}
We present a new method for the challenging source-free UDA problem. Without accessing source data or altering source training, our method achieves comparable or even better performance than source data-dependent approaches on cross-domain fundus image segmentation. 
The proposed pseudo label denoising schemes are also applicable to other scenarios such as semi-supervised learning.
One limitation of the work is that we consider the generally minor domain shift under which the pseudo labels can provide a meaningful supervision base. 
Actually such kind of domain shift commonly exits in datasets collected with different scanners, to which our method provides a practical solution. 
For the more severe cross-modality domain shift, the source network may largely under-segment objects on the target images, which may make the limited number of pseudo labels for positive classes difficult to achieve desired model self-adaptation. 
In future studies, we will further explore the source domain distribution information embedded in the source model or knowledge prior about the shape to jointly work with the pseudo-labeling for handling more severe domain shift.
\\
\\
\textbf{Acknowledgements.} This work was supported in part by Key-Area Research and Development Program of Guangdong Province, China (2020B010165004), National Natural Science Foundation of China with Project No. U1813204, and Hong Kong Innovation and Technology Fund (Project No. ITS/311/18FP and GHP/110/19SZ).

\bibliographystyle{splncs04.bst}

\bibliography{reference}
\end{document}